# Universal Silicon Microwave Photonic Spectral Shaper

Xin Guo,[1,2,3] Yang Liu,[1,3] Tangman Yin,[1,3] Blair Morrison,[1,3] Mattia Pagani,[1,3] Okky Daulay,[2] Wim Bogaerts,[4] Benjamin J. Eggleton,[1,3] Alvaro Casas-Bedoya,[1,3,**] and David Marpaung[2, *]

[1]*Institute of Photonics and Optical Science (IPOS), School of Physics, The University of Sydney, NSW 2006, Australia*
[2]*Nonlinear Nanophotonics group-LPNO, MESA+ Institute for Nanotechnology, University of Twente, Enschede, The Netherlands*
[3]*The University of Sydney Nano Institute (Sydney Nano), The University of Sydney, NSW 2006, Australia*
[4]*Ghent University - IMEC, Department of Information Technology, Photonics Research Group Technologiepark-Zwijnaarde 126, 9052 Gent, Belgium*
[*]*david.marpaung@utwente.nl*
[**]*alvaro.casasbedoya@sydney.edu.au*

**Abstract:** Optical modulation plays arguably the utmost important role in microwave photonic (MWP) systems. Precise synthesis of modulated optical spectra dictates virtually all aspects of MWP system quality including loss, noise figure, linearity, and the types of functionality that can be executed. But for such a critical function, the versatility to generate and transform analog optical modulation is severely lacking, blocking the pathways to truly unique MWP functions including ultra-linear links and low-loss high rejection filters. Here we demonstrate versatile RF photonic spectrum synthesis in an all-integrated silicon photonic circuit, enabling electrically-tailorable universal analog modulation transformation. We show a series of unprecedented RF filtering experiments through monolithic integration of the spectrum-synthesis circuit with a network of reconfigurable ring resonators.

## I. INTRODUCTION

In a canonical microwave photonic (MWP) system [1,2], radiofrequency (RF) signals are shaped and manipulated using optical techniques and components for wider processing bandwidth and advanced functionalities, which include tunable filtering [3,4], microwave beamsteering [5], tailored RF waveform generation [6], and RF spectrum analysis [7]. Optical modulation, as the step to translate an RF signal into the optical domain, is the most critical step in all of these systems with its significance going beyond simple encoding of the RF information onto the optical carrier. In the frequency domain, the optical modulation synthesizes a spectrum consisting of an optical carrier and two first-order RF-modulated sidebands. The phase and amplitude relations between these spectral components are essential in determining the type of functionalities that can be achieved in MWP systems. Upon photodetection, the mixing products of the optical carrier and the RF sidebands will interfere at the RF frequencies and dictate the phase and amplitude of the output RF signals.

Given the importance of optical modulation, it is rather surprising that the majority of MWP functionalities rely on just three types of modulation: phase modulation (PM), intensity modulation (IM), or single sideband modulation (SSB). This is in stark contrast with recent trends in the field where more and more programmable and general-purpose devices capable of synthesizing various MWP functions have been investigated [8-12]. Recent research results, however, have hinted that complete phase and amplitude control of an optical modulation spectrum can create opportunities for new and enhanced RF-photonic signal processing functions free from any performance trade-off, including advanced RF photonic filters with minimal loss and maximal rejection [3,13,14]. Removing these trade-offs will be critical for transitioning RF photonic systems to field-deployable technologies. But such a versatile spectrum synthesis function is presently unrealizable with currently available optical modulators which always create optical carrier and RF sidebands with interrelated phase and amplitudes. A device that targets an entirely new level of freedom in synthesizing a modulated spectrum will thus be significant to unlock unexplored degrees of freedom in microwave photonics.

## II. PRINCIPLE OF OPERATION

In this work we present a circuit to synthesize and shape arbitrary RF photonic spectra using an all-integrated silicon photonics chip. A modulated optical spectrum at the input of the circuit will be transformed into an output spectrum with spectral components (optical carrier and RF sidebands) entirely independent in phase and amplitude, essentially performing a universal modulation transformation. The operation of this circuit is markedly different compared to prior techniques specifically aimed at phase-to-intensity modulation (PM-IM) conversion [15,16]. Such techniques rely solely on the phase rotation of the optical carrier and thus can only switch between PM and IM.

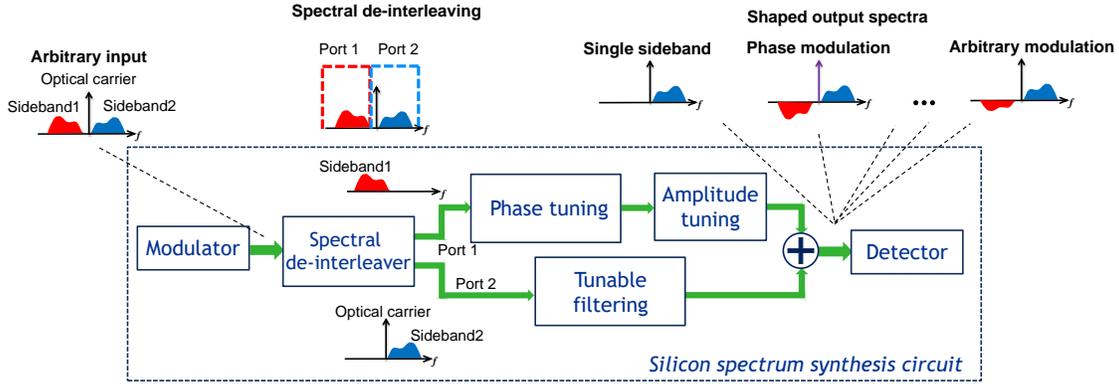

**Fig. 1.** Principle of operation of the spectrum synthesis circuit. The spectral de-interleaver is used to isolate a sideband from the rest of the RF-modulated optical spectrum. A cascade of tunable phase shifter and attenuator are used to manipulate phase and amplitude of the isolated sideband. Recombination with the unprocessed spectral components yields a versatile spectrum synthesis, producing various modulation formats such as single-sideband modulation, phase modulation, and advanced arbitrary modulation. Moreover, the combination of a tunable filtering element with the spectrum synthesis circuit leads to advanced reconfigurable RF photonic filtering functionalities.

The operational principle of the spectrum synthesis circuit is depicted in Fig. 1. The two output ports of spectral de-interleaver consist of completed complementary spectra. To achieve proper spectral synthesis and modulation transformation, one sideband of an input RF photonic spectrum is spatially isolated from the optical carrier and the remaining sideband via a spectral de-interleaver. The phase and amplitude of the isolated sideband can thus be further manipulated through an in-line phase shifter and a tunable attenuator, respectively. In principle, combining this tailored sideband with the rest of the spectrum will complete the modulation transformation process.

This versatile spectrum synthesis process can further be combined with complementary signal processing intended for activating RF photonic functionalities. For example, a tunable filter network can be placed in the path of the optical carrier and the un-isolated sideband prior to recombination. Proper tailoring of the RF modulated spectrum in combination with complex optical filtering provided by the filter network will lead to advanced high-performance RF filters at the photodetector output. The operation of the spectrum synthesis circuit can be verified through inspection of optical spectra at various tapped output ports prior to photodetection, as well as measuring the RF mixing products of the optical carrier and the two sidebands at the photodetector output.

## III. RESULTS

### A. Circuit implementation

The implementation of the spectrum synthesis circuit is illustrated in Fig. 2. We realize the spectral de-interleaver using a Mach-Zehnder interferometer (MZI) loaded with three ring resonators (MZI+3 rings) topology [10]. The two outputs of the de-interleaver exhibit square-shaped, flat-top complementary filters. The de-interleaver outputs containing the isolated sideband is routed to a cascade of a thermo-optic phase shifter and a tunable coupler implemented as a balanced MZI, for independent phase and amplitude tailoring. The complementary output of the de-interleaver carries the optical carrier and the un-isolated sideband, and is routed to a cascade of an all-pass and an add-drop ring resonators. These rings are used to implement tunable filtering on the un-isolated sideband. The two waveguide paths are then re-combined in a 3-dB coupler to synthesize a new spectrum with designer phase and amplitude relations between its spectral components. One of the two outputs of the 3-dB coupler are sent to an on-chip photodetector and the other routed off-chip through a grating coupler.

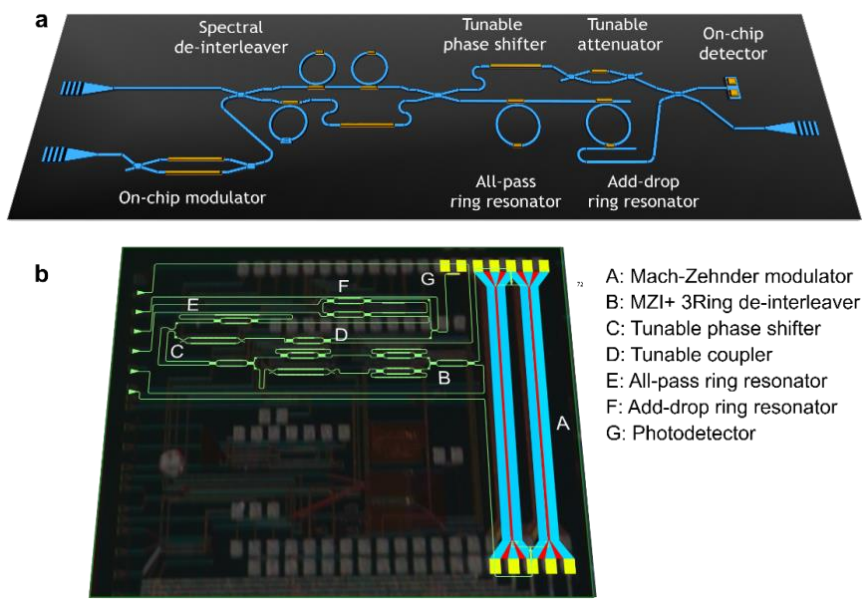

**Fig. 2.** Implementation of the spectrum synthesis circuit in a silicon chip. (a) Artistic representation of the silicon spectrum synthesis circuit showing the active components (a modulator and a photodetector) and passive optical circuit containing the Mach-Zehnder interferometer loaded with 3 ring resonators (MZI+3rings) as the spectral de-interleaver. A cascade of all-pass and add-drop ring resonators are used to implement RF photonic filtering; (b) The circuit layout of the silicon chip overlaid on the photograph of the chip.

The spectrum synthesis circuit was fabricated in IMEC's silicon photonic iSiPP50G process through the Europractice multi-project wafer service. All key components in the circuit including the phase shifters, tunable couplers, and the de-interleaver were fully tunable through thermo-optic tuning. The spectrum synthesis circuit was designed to flexibly work either with an external modulator or with an-on-chip Mach-Zehnder modulator (see Appendix A for the details of the device design and fabrication). Fig. 2(c) shows a picture of our silicon chip.

We characterized the individual components in the circuit and compared the measurement results with our model, developed using z-transform approach, as described in Supplement 1 section 1. The transmission spectrum of the de-interleaver, is depicted in Fig. 3(a). Optimized tuning of the heaters on the circuit leads to a de-interleaver with complementary outputs with 30 GHz-wide pass and stop bands and a rejection of more than 20 dB. The filtering response of all-pass ring resonator and add-drop ring resonator of the followed ring network circuits are shown in Fig. 3(b) and 3(c), respectively. The free-spectral range (FSR) of the rings are 0.4 nm (50 GHz). Comparing the measured responses with simulation, we extract the maximum loaded Q-factors of the all-pass and add-drop rings to be 120,000. At critical coupling, the Q-factor and finesse of the all-pass ring were 68,000 and 17.6, respectively (Fig. 3(b)).

## B. Modulation transformation experiments

To demonstrate the versatility of our approach, we controlled our circuit to synthesize various RF modulation schemes from a conventional phase or intensity modulation input. In these series of experiments, described in details in Appendix B, we used the on-chip photodetector in combination with external electro-optic modulators (as illustrated Fig. 4(a)) to ensure high bandwidth operation. The on-chip modulator, on the other hand, exhibited limited bandwidth of (11 GHz) as compared to the de-interleaver bandwidth (30 GHz). Moreover, we observed high internal RF crosstalk in our chip (See Supplement 1 section 4 for details).

In the first experiment an intensity modulated signal is sent to the circuit and the optical phase shifter is used to modify the phase of the isolated sideband to synthesize phase modulation from an intensity modulation input (IM-PM conversion). When both sidebands are in-phase, the measured RF transmission is high, as shown in the blue trace of Fig. 4(b). We then tuned the circuit to rotate the phase of the isolated sideband and synthesize phase modulation. When the sidebands are out-of-phase the RF transmission is minimized, indicating phase modulation (red trace in Fig. 4(b)). The extinction of the IM-PM conversion was 15 dB. We then replaced the intensity modulator with a phase modulator and repeated the experiments where results depicted in Fig. 4(c). In this experiment, phase modulation input was transformed to intensity modulation output. For both experiments we achieved modulation transformation of 10 GHz bandwidth. An RF transmission extinction up to 15 dB can be achieved in these experiments solely by electric tuning the phase shifter after the bar port of the de-interleaver. The bandwidth and extinction are mainly limited by the roll-off and the dispersion of the MZI+3 rings de-interleaver notably at the transition band. The details of the simulations and the parameters of the experiments can be found in Supplement 1 section 2.

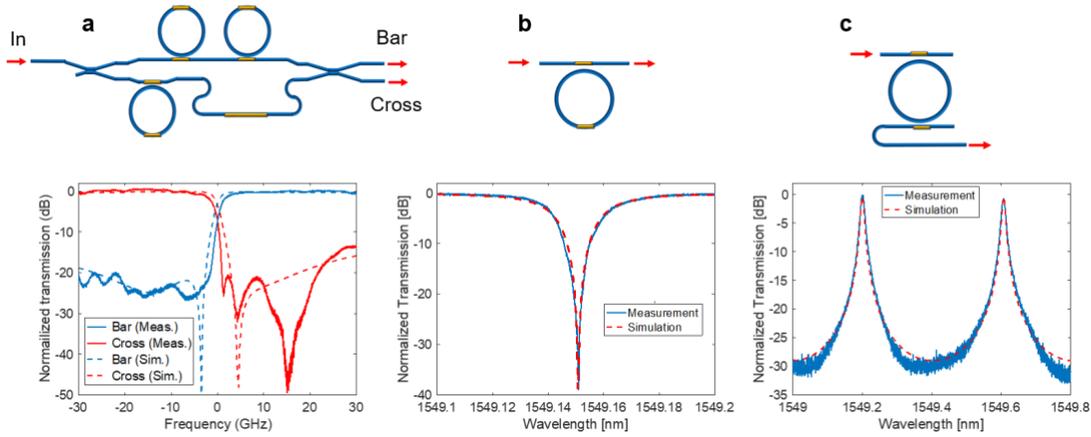

**Fig. 3.** Characterization of the individual components in the circuit. (a) Transmission spectrum of the de-interleaver revealing 30 GHz channel bandwidth and more than 20 dB extinction; (b) Measured (solid line) and modelled (dashed line) filtering response of an all-pass ring; (c) Measured (solid line) and simulated (dashed line) filtering response of an add-drop ring at the drop port.

We verified that the modulation transformation was achieved purely from phase tuning effects by inspecting the optical spectra of the corresponding RF traces. Both traces in Fig. 4(b), either with high and low RF transmissions, were showing practically identical optical spectra as shown in Fig. 4(d). It should be noted that in the experiment, the central frequency of the laser needs to be slightly adjusted due to the drift in central frequencies of the de-interleaver passbands, which was attributed to the thermal crosstalk in the circuit. With further characterization, this could be compensated by adaptive control of the thermal tuners [17].

## C. Advanced RF filtering experiments

Reconfigurable modulation transformations will enable various RF photonic filtering schemes in a single chip-scale device using the simplest modulator configuration such as a phase modulator, which prior to this work would have been impossible especially since silicon photonics modulators are known to exhibit not just phase modulation but also spurious amplitude modulation. Here, we demonstrated three distinct RF photonic filtering topologies that would require very different kinds of input modulations to the photonic filtering elements. Using our silicon photonic circuit, we shaped RF photonic spectra from a common intensity modulator input to match the subsequent filtering function provided by on-chip resonator networks and show conventional SSB bandpass and bandstop filters as well as a high-extinction RF-interference notch filter in a single reconfigurable device, which prior to this work would have been impossible.

Fig. 5 depicts the experimental schematic and the optical or RF spectrum at each stage of the three RF photonic filter topologies demonstrated here. We first configured the modulation transformer to synthesize single sideband (SSB) modulation by activating the tunable coupler to completely block the isolated sideband (Fig. 5 (a)). We sent the remaining carrier and sideband to an all-pass ring resonator to form an SSB RF photonic notch filter. The resulting RF filter response is shown as a dashed-line in Fig. 5(d). The rejection of the RF notch filter is proportional to rejection in optical domain, which was set to 7 dB.

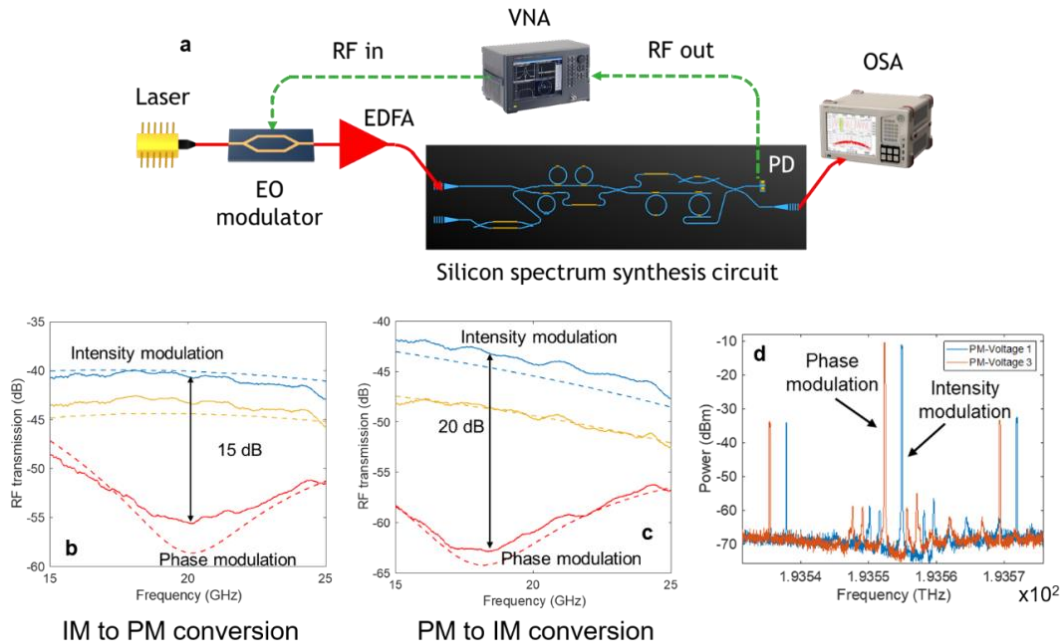

**Fig. 4.** Setup and results of modulation transformation experiments. (a) Schematic of the experiment setup used to characterise RF modulation transformation and RF photonic filtering. EDFA: erbium-doped fiber amplifier, PD: photodetector, VNA: RF vector network analyzer, OSA: optical spectrum analyzer; (b) Results of intensity to phase modulation (IM to PM) conversion experiment showing 15 dB extinction, in which the intensity modulation produces a high RF link gain, while the phase modulation a significantly reduced RF link gain due to the destructive interference of out-of-phase sidebands; (c) Results of phase to intensity modulation (PM to IM) conversion experiment exhibiting 20 dB extinction; (d) Synthesized optical spectrum with intensity and phase modulation reveal similar features in the amplitude spectrum, proving that the modulation format transformation was due to phase rotation of the sideband.

Next, we reconfigured the circuit to show a complex phase cancellation filter [18] that can amplify the shallow rejection of the all-pass ring to a much higher extinction. This requires the modulation transformer to synthesize a new spectrum with unequal-amplitude and anti-phase sidebands (Fig. 5(b)). Here the isolated sideband was attenuated by 7 dB and rotated in phase to exhibit $\pi$ phase shift with respect to the un-isolated sideband. We then sent this spectrum to the all-pass ring. At the notch frequency, the amplitude of the two sidebands are equal but they are opposite in phase which, upon photodetection, will form an ultra-high rejection RF notch filter of 38 dB (solid line of Fig. 5(d)). Data for the central frequency tuning of the notch filters can be found in Supplement 1 section 5.

In the third filtering experiment, we demonstrate bandpass filtering using an add-drop ring resonator (Fig. 5c). The challenge in this filtering is that the passband of the ring is too narrow to pass both the optical carrier and a tunable-frequency sideband, so forming a tunable RF photonic bandpass filter without precise carrier re-insertion is, in fact, impossible. We mitigate this limitation using our modulation transformer. First we synthesize an SSB + carrier spectrum at the input. We configured the modulation transformer to separate the carrier and the sideband, where the latter is being passed to the add-drop ring to be filtered. The 3-dB coupler re-combined the carrier and the processed sideband to form a true SSB RF photonic bandpass filter. We then tuned the resonance frequency of the add-drop ring, resulting in a tuned central frequency of the RF bandpass filter, as depicted in Fig. 5e.

## IV. DISCUSSION

The concept of RF photonic spectrum synthesis presented here can be extended beyond modulation transformation and filtering. Full phase and amplitude control of optical carrier and RF sidebands up to higher order can be used to linearize RF photonic links [19-21], to create tailorable RF waveforms [6,22], to control RF photonic delay lines [14,23], to develop beamforming and photonics radar system [24,25], and to improve the gain and noise figure of a microwave photonic systems [18,26,27]. Independent control of optical carrier and sideband amplitudes in the circuit can be used to optimize the modulation index and power ratio of these spectral components and to emulate low biasing of a Mach-Zehnder modulator. These techniques will lead to reduction of the noise power spectral densities and noise figure of the system.

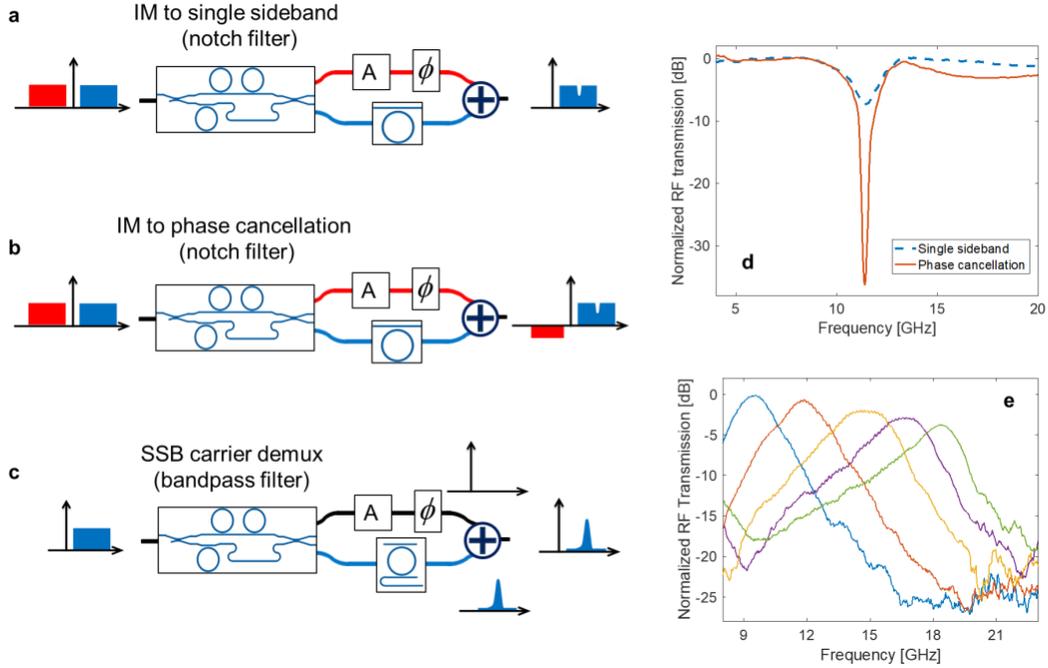

**Fig. 5.** Configurations and results of RF filtering experiments. (a) Single sideband notch filter experiment where the spectrum synthesis circuit was used to remove the lower RF sideband; (b) Phase cancellation notch filter, where the circuit was used to synthesize RF sidebands with unbalanced amplitude and opposite phase; (c) A bandpass filter where the circuit was used as a carrier re-insertion network; (d) Comparison of RF notch filter responses from the single-sideband scheme in (a) and the phase cancellation filter in (b). With the same optical filter response, the single-sideband filter shows a shallow response (blue dashed-trace), while the phase cancellation filter exhibits a deeper rejection enhanced by 30 dB (red solid trace); (e) Measured central frequency-tuned RF responses of the bandpass filter in shown (c).

Versatile tailoring of RF photonic modulated spectrum is a critical function in the new paradigm of field programmable and general-purpose integrated photonics [8-10,28] and is an important building block to realize cascadable on-chip RF photonic functions where each function, for example delay line followed by RF photonic filtering, can only be optimized with particular modulation format [29]. Insertion of modulation transformation in between these functions can unlock the potential of cascaded on-chip RF photonics with minimized E/O and O/E conversion losses.

The work presented here is a fundamental demonstration of RF photonic spectrum synthesis and modulation transformation principles, but it is currently limited by realizable technologies. Two essential building blocks to execute modulation transformation proposed here, namely the spatial separation of individual spectral components and the independent phase and amplitude tailoring are still beset by non-idealities in the complex transfer functions of the de-interleaver, the phase shifter, and the tunable coupler. For example, the roll-off in the magnitude response of the de-interleaver limits the lowest RF frequencies that can be processed. This roll-off is a very sensitive function of the waveguide propagation loss, which amounts to 1.2 dB/cm in our demonstration. Shifting these function to lower loss platforms such as silicon nitride will lead to considerable improvement and filter sharpness. On the other hand, the nonlinear phase response at the transition bands of the de-interleaver restricts the achievable modulation transformation bandwidth. This is a more fundamental limitation that cannot simply be solved by reducing waveguide losses. Alternative de-interleaver design with more linear phase response, such as using a cascaded Mach-Zender interferometers (MZIs) [30] can be considered.

Broadband and independent phase and amplitude tailoring is essential to achieve tractable modulation transformation. Present implementation with tunable phase shifter and MZI-based tunable coupler leads to limited bandwidth and interrelated phase and amplitude tuning. When tuned, the tunable coupler exhibits parasitic phase rotation that always needs to be compensated by tuning the phase shifter (see Supplement 1 section 1). Bandwidth limitation of spectrum transformation also comes from optical path-length difference between the de-interleaved spectra. A thermo-optic phase shifter is not suitable to compensate this difference and instead a tunable optical delay line is required.

## V. CONCLUSION

In summary, we demonstrated a universal RF photonic spectral shaper with both passive and active components on a monolithic silicon photonic circuit. The circuit allows for independent tailoring of phase and amplitude of the optical carrier and RF sidebands leading to the first demonstration of universal RF photonic modulation transformation. With this entirely novel concept, we demonstrated three distinct kinds of RF photonics filters using a simple intensity modulator, which was previously unattainable using any other photonic circuits. This work will serve as the basis of further RF photonic noise, dynamic range, and filter optimizations and stabilization through modulation spectral shaping and transformation.

## APPENDIX A: DETAILS OF DEVICE DESIGN AND FABRICATION

The silicon chip was designed with the IPKISS parametric design framework [31] and fabricated in IMEC's iSiPP50G process [32] through the Europractice multi-project wafer service. The key building blocks are implemented as SOI nanowires with 220 nm × 450 nm dimension. The optical waveguide loss is 1.2 dB/cm. All components are tunable using thermo-optic tuning. To get the device with full reconfiguration capability and to compensate for any fabrication imperfection, we designed the 3-dB couplers of the de-interleaver and the waveguide-ring couplers as a tunable symmetric MZI coupler. The modulator is a 1.5mm long Mach-Zehnder modulator with depletion-type phase shifters with 11 GHz RF bandwidth and the photodetector is a 40 GHz Ge-PD with responsivity of 0.8 A/W, both from the provided iSiPP50G design kit. The de-interleaver was designed to have a passband width of 30 GHz and peak rejection of 30 GHz. The all-pass and add-drop rings both have a free spectral range (FSR) of 50 GHz. The maximum quality factor of the all-pass and the add-drop rings was 120,000. The silicon chip was wire bonded locally onto a bespoke PCB for ease of experiments. The optical signal is coupled in and out of the chip using grating couplers, with a 3-dB loss per coupler.

## APPENDIX B: DETAILS OF THE EXPERIMENTAL SETUP

In the experiments we used a tunable diode laser (Yenista TUNICS T100S-HP) with 0 dBm of output power, connected to an intensity modulator (Thorlabs, LN05S-FC) or a phase modulator (Thorlabs, LN27S-FC). The output of modulator is connected to an EDFA (Amonics) with maximum output power of 19.1 dBm. The chip is mounted on a vertically-coupled computer controlled fiber stage (Maple Leaf Photonics), the fiber array contains 19 output ports with an interval of 127 microns. The light being coupled out is sent to a photodetector (Finisar, XPDV2120RA-VF-FP) or an optical spectrum analyzer (Apex, AP2083A). For the RF transmission measurement, a frequency-swept RF signal with 0 dBm power was supplied from and measured on a 10 MHz-43.5 GHz vector network analyzer (Agilent, PNA 5224A).

## FUNDING

Australian Research Council (ARC) through ARC DECRA (DE150101535), Laureate Fellowship (FL120100029), and Center of Excellence CUDOS (CE110001018).

Netherlands Organisation for Scientific Research NWO Vidi (15702) and Start Up (740.018.021)

**ACKNOWLEDGMENTS**

The authors acknowledge the facilities as well as the scientific and technical assistance of the Research & Prototype Foundry Core Research Facility at the University of Sydney, part of the Australian National Fabrication Facility. We acknowledge technical assistance from C. Cantaloube and use of facilities provided by Microsoft Quantum Sydney.

**DATA AVAILABILITY**

The data that support the findings of this study are available from the corresponding author upon reasonable request.

**REFERENCES**

1. J. Capmany, and D. Novak, "Microwave photonics combines two worlds," Nat. Photon. **1,** 319-330 (2007).
2. D. Marpaung, J. Yao, and J. Capmany, "Integrated microwave photonics," Nat. Photon. **23,** 80-90 (2019).
3. D. Marpaung, B. Morrison, M. Pagani, R. Pant, D. Choi, B. Luther-Davies, S. J. Madden, and B. J. Eggleton, "Low-power, chip-based stimulated Brillouin scattering microwave photonic filter with ultrahigh selectivity," Optica **2,** 76-83 (2015).
4. X. Xu, M. Tan, J. Wu, T. G. Nguyen, S. T. Chu, B. E. Little, R. Morandotti, A. Mitchell, and D. J. Moss, "High performance RF filters via bandwidth scaling with Kerr micro-combs," APL Photonics **4,** 026102 (2019).
5. V. C. Duarte, J. G. Prata, C. F. Ribeiro, R. N. Nogueira, G. Winzer, L. Zimmermann, R. Walker, S. Clements, M. Filipowicz, M. Napierała, and T. Nasiłowski, "Modular coherent photonic-aided payload receiver for communications satellites," Nat. Commun. **10,** 1-9 (2019).
6. J. Wang, H. Shen, L. Fan, R. Wu, B. Niu, L. T. Varghese, Y. Xuan, D. E. Leaird, X. Wang, F. Gan, and A. M. Weiner, "Reconfigurable radio-frequency arbitrary waveforms synthesized in a silicon photonic chip," Nat. Commun. **6,** 1-8 (2015).
7. B. Maurizio, X. Wang, M. Li, L. Chrostowski, and J. Azaña, "Wideband dynamic microwave frequency identification system using a low-power ultracompact silicon photonic chip," Nat. Commun. **7,** 13004 (2016).
8. W. Liu, M. Li, R. S. Guzzon, E. J. Norberg, J. S. Parker, M. Lu, L. A. Coldren, and J. Yao, "A fully reconfigurable photonic integrated signal processor," Nat. Photon. **10,** 190-195 (2016).
9. J. Capmany, I. Gasulla, and D. Pérez, "Microwave photonics: the programmable processor," Nat. Photon. **10,** 6–8 (2016).
10. P. Daniel, I. Gasulla, L. Crudgington, D. J. Thomson, A. Z. Khokhar, K. Li, W. Cao, G. Z. Mashanovich, and J. Capmany, "Multipurpose silicon photonics signal processor core," Nat. Commun. **8,** 1-9 (2017).
11. J. Liu, E. Lucas, A. S. Raja, J. He, J. Riemensberger, R. N. Wang, M. Karpov, H. Guo, R. Bouchand, and T. J. Kippenberg, "Photonic microwave generation in the X-and K-band using integrated soliton microcombs," Nat. Photon. 1-6 (2020).
12. J. Hulme, M. J. Kennedy, R. Chao, L. Liang, T. Komljenovic, J. Shi, B. Szafraniec, D. Baney, and J. E. Bowers, "Fully integrated microwave frequency synthesizer on heterogeneous silicon-III/V," Opt. Express **25,** 2422-2431 (2017).
13. A. Casas-Bedoya, B. Morrison, M. Pagani, D. Marpaung, and B. J. Eggleton, "Tunable narrowband microwave photonic filter created by stimulated Brillouin scattering from a silicon nanowire," Opt. Lett. **40,** 4154-4157 (2015).
14. L. Zhuang, M. Burla, C. Taddei, C. GH Roeloffzen, M. Hoekman, A. Leinse, K. J. Boller, and A. J. Lowery, "Integrated microwave photonic splitter with reconfigurable amplitude, phase, and delay offsets," Opt. Lett. **40,** 5618-5621 (2015).
15. L. Zhuang, M. Hoekman, C. Taddei, A. Leinse, R. G. Heideman, A. Hulzinga, J. Verpoorte, R. M. Oldenbeuving, P. W. van Dijk, K. J. Boller, and C. G. Roeloffzen, "On-chip microwave photonic beamformer circuits operating with phase modulation and direct detection," Opt. Express **22,** 17079-17091 (2014).
16. W. Li, N. Zhu, L. Wang, and H. Wang, "Broadband phase-to-intensity modulation conversion for microwave photonics processing using Brillouin-assisted carrier phase shift," J. Light. Technol. **29,** 3616-3621 (2011).
17. M. Milanizadeh, S. Ahmadi, M. Petrini, D. Aguiar, R. Mazzanti, F. Zanetto, E. Guglielmi, M. Sampietro, F. Morichetti, and A. Melloni, "Control and Calibration Recipes for Photonic Integrated Circuits," IEEE J. Sel. Top. Quantum Electron. **26,** 1-10 (2020).
18. D. Marpaung, B. Morrison, R. Pant, C. Roeloffzen, A. Leinse, M. Hoekman, R. Heideman, and B. J. Eggleton. "$Si_3N_4$ ring resonator-based microwave photonic notch filter with an ultrahigh peak rejection," Opt. Express **21,** 23286-23294 (2013).
19. R. Wu, T. Jiang, S. Yu, J. Shang, and W. Gu, "Multi-Order Nonlinear Distortions Analysis and Suppression in Phase Modulation Microwave Photonics Link," J. Light. Technol. **37,** 5973-5981 (2019).
20. D. Zhu, J. Chen, and S. Pan, "Linearized phase-modulated analog photonic link with the dispersion-induced power fading effect suppressed based on optical carrier band processing," Opt. Express **25,** 10397-10404 (2017).

21. Z. Xie, Zhipeng, S. Yu, S. Cai, and W. Gu, "Simultaneous Improvements of Gain and Linearity in Dispersion-Tolerant Phase-Modulated Analog Photonic Link," IEEE Photon. J. **9,** 1-12 (2017).
22. Z. Zhu, Y. Liu, M. Merklein, Z. Zhang, D. Marpaung, and B. J. Eggleton, "$Si_3N_4$-chip-based versatile photonic RF waveform generator with a wide tuning range of repetition rate," Opt. Lett. **45,** 1370-1373 (2020).
23. Y. Liu, A. Choudhary, D. Marpaung, and B. J. Eggleton, "Gigahertz optical tuning of an on-chip radio frequency photonic delay line," Optica **4,** 418-423 (2017).
24. X. Xue, Y. Xuan, C. Bao, S. Li, X. Zheng, B. Zhou, M. Qi, and A. M. Weiner, "Microcomb-based true-time-delay network for microwave beamforming with arbitrary beam pattern control," J. Light. Technol. **36,** 2312-2321 (2018).
25. P. Ghelfi, F. Laghezza, F. Scotti, G. Serafino, A. Capria, S. Pinna, D. Onori, C. Porzi, M. Scaffardi, A. Malacarne, and V. Vercesi, "A fully photonics-based coherent radar system," Nature **507,** 341-345 (2014).
26. Y. Liu, D. Marpaung, A. Choudhary, and B. J. Eggleton, "Lossless and high-resolution RF photonic notch filter," Opt. Lett. **41,** 5306-5309 (2016).
27. Z. Zhu, Y. Liu, M. Merklein, O. Daulay, D. Marpaung, and B, J. Eggleton, "Positive link gain microwave photonic bandpass filter using $Si_3N_4$-ring-enabled sideband filtering and carrier suppression," Opt. Express **27,** 31727-31740 (2019).
28. J. Feldmann, N. Youngblood, C. D. Wright, H. Bhaskaran, and W. H. P. Pernice, "All-optical spiking neurosynaptic networks with self-learning capabilities," Nature **569,** 208-214 (2019).
29. S. X. Chew, L. Nguyen, X. Yi, S. Song, L. Li, P. Bian, and R. Minasian, "Distributed optical signal processing for microwave photonics subsystems," Opt. Express **24,** 4730-4739 (2016).
30. F. Horst, W. M.J. Green, S. Assefa, S. M. Shank, Y. A. Vlasov, and B. J. Offrein, "Cascaded Mach-Zehnder wavelength filters in silicon photonics for low loss and flat pass-band WDM (de-) multiplexing," Opt. Express **21**, 11652-11658 (2013)
31. W. Bogaerts, M. Fiers, M. Sivilotti, and P. Dumon, "The IPKISS photonic design framework," in Optical Fiber Communications Conference and Exposition (OFC). 20-24 March 2016 (OSA, 2016).
32. M. Pantouvaki, S. A. Srinivasan, Y. Ban, P. D. Heyn, P. Verheyen, G. Lepage, H. Chen, J. D. Coster, N. Golshani, S. Balakrishnan, and P. Absil, "Active components for 50 Gb/s NRZ-OOK optical interconnects in a silicon photonics platform," J. Light. Technol. **35,** 631-638(2017).

# Supplementary Information: Universal Silicon Microwave Spectral Shaper

**Xin Guo,**[1,2,3] **Yang Liu,**[1,3] **Tangman Yin,**[1,3] **Blair Morrison,**[1,3] **Mattia Pagani,**[1,3] **Okky Daulay,**[2] **Wim Bogaerts,**[4] **Benjamin J. Eggleton,**[1,3] **Alvaro Casas-Bedoya,**[1,3,**] **and David Marpaung**[2, *]

[1]Institute of Photonics and Optical Science (IPOS), School of Physics, The University of Sydney, NSW 2006, Australia
[2]Nonlinear Nanophotonics group-LPNO, MESA+ Institute for Nanotechnology, University of Twente, Enschede, The Netherlands
[3]The University of Sydney Nano Institute (Sydney Nano), The University of Sydney, NSW 2006, Australia
[4]Ghent University - IMEC, Department of Information Technology, Photonics Research Group Technologiepark-Zwijnaarde 126, 9052 Gent, Belgium
[*]david.marpaung@utwente.nl
[**]alvaro.casasbedoya@sydney.edu.au

## 1a. Transfer function of components in the spectrum synthesis circuit

We list all the passive building blocks and their transfer function here. The entire model of the spectrum synthesis circuit can be accessed by assigning each building block with specific transfer function and cascading them.

| Structure | Layout | Transfer function |
|---|---|---|
| Bus waveguide | 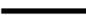 | $H_{bw}(z) = \gamma z^{-1}$, where $z^{-1} = \exp\left(-j \cdot \frac{2\pi n_g L}{\lambda}\right)$ and $\gamma = \exp(-\alpha L)$ |
| Phase shifter | 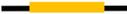 | $H_{ps} = e^{-j\varphi}$ |
| 3-dB coupler | 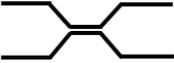 | $H_{dc} = \begin{bmatrix} \sqrt{0.5} & -j\sqrt{0.5} \\ -j\sqrt{0.5} & \sqrt{0.5} \end{bmatrix}$ |
| Tunable coupler | 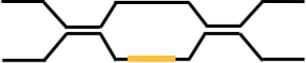 | $H_{tc} = \begin{bmatrix} \sqrt{0.5} & -j\sqrt{0.5} \\ -j\sqrt{0.5} & \sqrt{0.5} \end{bmatrix} \cdot \begin{bmatrix} 1 \\ e^{-j\varphi} \end{bmatrix} \cdot \begin{bmatrix} \sqrt{0.5} & -j\sqrt{0.5} \\ -j\sqrt{0.5} & \sqrt{0.5} \end{bmatrix}$ |
| Ring resonator | 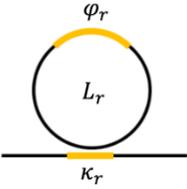 | $H_r(z) = \dfrac{c_r - a_r \gamma_r z_r^{-1}}{1 - c_r a_r \gamma_r z_r^{-1}}$ |
| MZI +3 rings de-interleaver | 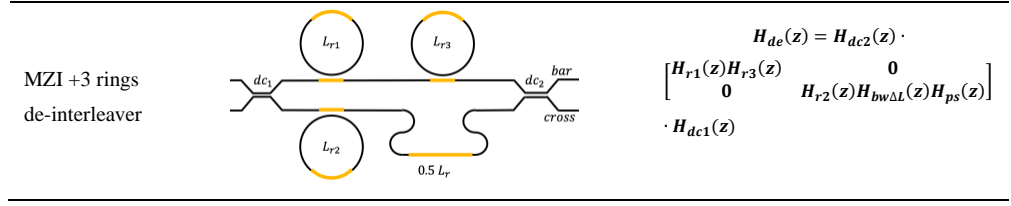 | $H_{de}(z) = H_{dc2}(z) \cdot$ $\begin{bmatrix} H_{r1}(z)H_{r3}(z) & 0 \\ 0 & H_{r2}(z)H_{bw\Delta L}(z)H_{ps}(z) \end{bmatrix}$ $\cdot H_{dc1}(z)$ |

Table S1. Building blocks of modulation transformer and their transfer functions.

The transfer functions of building blocks are shown above in Table S1, where $j$ is the imaginary unit, L is the length of the waveguide, $n_g$ is the group index and $\gamma$ represents the loss. The phase shifter is modelled as a tunable phase delay $\varphi$ to the waveguide. The tunable coupler can be regarded as two cascaded 3-dB couplers, or a balanced MZI. The phase shifter set on one of the arms controls the power splitting ratio. For the ring resonator, $c_r = (1 - \kappa_r)^{1/2}$ where $\kappa$ stands for the power coupling coefficient; $a_r = e^{-j\varphi_L}$ and $\gamma_r = \exp(-\alpha L_r)$, which are the phase and loss of the ring resonator. The transfer function of MZI+3 rings de-interleaver is assembled by the above building blocks. $H_{de}(z)$ is a $1 \times 2$ matrix that contains the transfer functions for bar port and cross port respectively.

### 1b. Modelling of tunable couplers

In this work, the tunable coupler was intended to give a pure amplitude tuning for tailoring the sideband amplitudes. In practice the tunable couplers were implemented as balanced MZI with a phase shifter in one of its arms. This implementation leads to a tunable coupler that exhibits a parasitic phase rotation which influence the desired phase setting. This unwanted phase rotation originates from the fact that once the heater is being tuned, the optical length of that arm will be changed which results in different interference (splitting ratio between two outputs ports), but the modified optical length also changes the absolute phase of the light wave, which always need to be compensated by tuning the phase shifter.

Fig. S1 shows, when the heater is being thermally tuned, the power splitting ratio and corresponding induced phase change of both ports of the tunable coupler. This effect has been taken into account in the modelling of the entire RF photonic system (see Supplementary 2b).

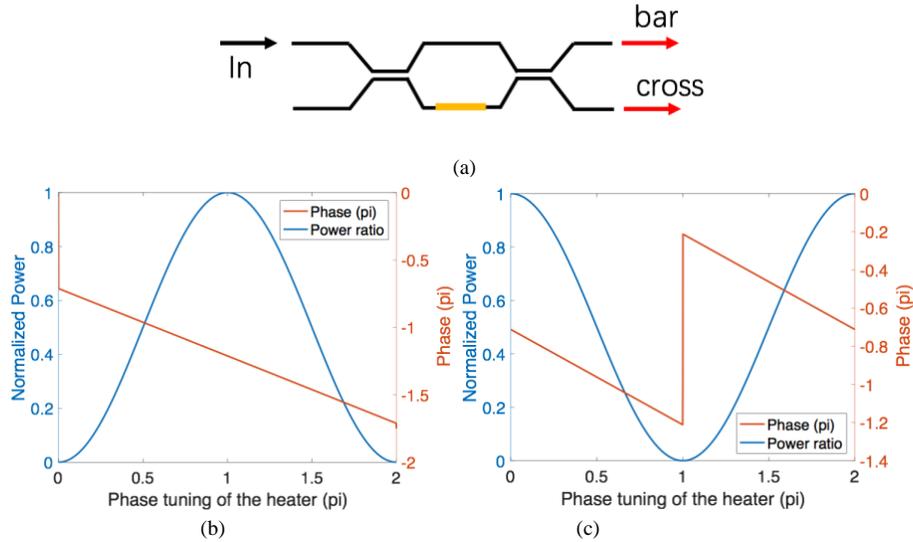

Fig. S1 (a) Schematic of a tunable coupler. Power and phase response of tunable coupler outputs in one complete tuning range from 0 to 2π: bar port (b) and cross port (c).

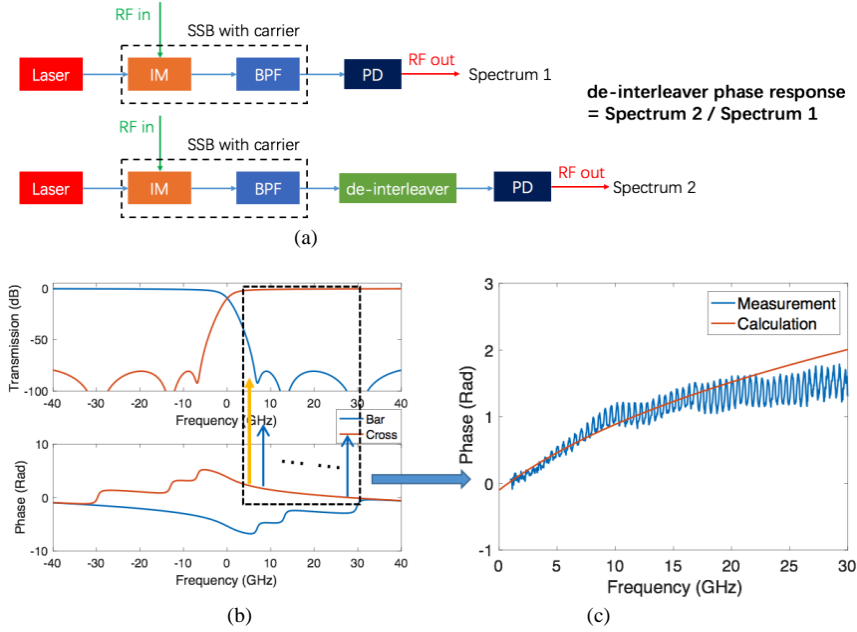

Fig. S2 (a) Experimental setup for the de-interleaver phase measurement; (b) Intensity and phase response of the de-interleaver; (c) Measured interleaver phase response. IM: intensity modulator; BPF: Bandpass filter; PD: Photodetector.

## 1c. Modelling of MZI+3Rings de-interleaver

Ideally, the signal separation device (de-interleaver) exhibits a rectangular amplitude response with a flat phase response over the frequency bands. However, each optical filter will induce their phase response in to the system. The dispersion of our de-interleaver results in the limited modulation transformation bandwith as shown in Fig. 4 in the paper.

To entirely understand the complex filtering response of the de-interleaver, we perform both theoretical and experimental investigations. Fig. S2 (a) shows the method of phase measurement. Fig. S2 (b) depicts the simulated intensity and phase response of the de-interleaver based on the transform function mentioned in Supplementary 1a. There is a linear phase response regime in the central passband, while its slope becomes sharper at the edge of the filter response. In the operation of the modulation transformer, the carrier was placed in the transition band of the filter.

To measure the phase response, we sent a single sideband signal (SSB), generated using bandpass-filtered intensity modulated signal, to the de-interleaver and detect the beat product of the optical carrier and the sideband in the PD and the VNA. As shown in Fig. S2 (a), we use the photodetector to measure the SSB signal first and record it as the base (Spectrum 1), then we let the SSB signal go through the de-interleaver and measure its response (Spectrum 2), the laser carrier was put into the starting point of passband (dashed box in Fig. S2 (b)). Because the phase response is relative, a division of Spectrum 2 and Spectrum 1 can be used to extract the phase response of the de-interleaver (Blue curve in Fig. S2 (c)).

## 2a. Derivation of photodetector output

We consider an optical signal that constitutes of carrier and two sidebands, which can be written as

**Eq. (S1)** $$A = A_- + A_c + A_+.$$

Where $A_-$, $A_c$ and $A_+$ stand for the fields of lower sideband, carrier and upper sideband. These three fields can be described as

**Eq. (S2)** $$A_- = E_- \cdot e^{j[(\omega_c - \omega_{RF})t + \varphi_-]}$$
**Eq. (S3)** $$A_c = E_c \cdot e^{j[\omega_c t + \varphi_c]}$$
**Eq. (S4)** $$A_+ = E_+ \cdot e^{j[(\omega_c + \omega_{RF})t + \varphi_+]}.$$

And the response of the photodetector is given by

**Eq. (S5)** $$I_{det} = \gamma_{PD} P_{out} \propto \gamma_{PD} AA^* = \gamma_{PD}(A_- + A_c + A_+)(A_- + A_c + A_+)^*,$$

where $\gamma_{PD}$ is the responsivity of the photodetector. Ignoring the second-order harmonics terms at $2\omega_{RF}$, the detected photocurrent can be expressed by

**Eq. (S6)** $$I_{det\_RF} \propto \gamma_{PD}[A_c(A_- + A_+)^* + A_c^*(A_- + A_+)].$$

Eq. 6 can be modified as

**Eq. (S7)** $$I_{det\_RF} \propto \gamma_{PD} \cdot 2Re[A_c(A_- + A_+)^*].$$

where

$$Re[A_c(A_- + A_+)^*] = Re\{E_c \cdot e^{j[\omega_c t + \varphi_c]} \cdot \{E_- \cdot e^{j[(\omega_c - \omega_{RF})t + \varphi_-]} + E_+ \cdot e^{-j[(\omega_c + \omega_{RF})t + \varphi_+]}\}\}$$
$$= Re\{E_c E_- \cdot e^{j[\omega_{RF} t + (\varphi_c - \varphi_-)]} + E_c E_+ \cdot e^{-j[\omega_{RF} t + (\varphi_+ - \varphi_c)]}\}$$
**Eq. (S8)** $$= E_c E_- \cos(\omega_{RF} t + \varphi_c - \varphi_-) + E_c E_+ \cos(\omega_{RF} t + \varphi_+ - \varphi_c).$$

The terms $\varphi_c - \varphi_-$ and $\varphi_+ - \varphi_c$ literally symbolize the phase difference between the carrier and the two sidebands, and the positive or negative sign does not matter. In that case, they can be written as $\Delta\varphi_-$ and $\Delta\varphi_+$. By substituting Eq. S8 in to Eq. S7, we have

**Eq. (S9)** $$I_{det\_RF} \propto 2\gamma_{PD}[E_c E_- \cos(\omega_{RF} t + \Delta\varphi_-) + E_c E_+ \cos(\omega_{RF} t + \Delta\varphi_+)]$$

For intensity modulation ($\Delta\varphi_- = \Delta\varphi_+$), Eq. S9 can be expressed by

**Eq. (S10)** $$I_{det\_RF} \propto 2\gamma_{PD}[E_c E_- \cos(\omega_{RF} t + \Delta\varphi_-) + E_c E_- \cos(\omega_{RF} t + \Delta\varphi_-)]$$
$$= I_{det\_RF} \propto 4\gamma_{PD} E_c E_- \cos(\omega_{RF} t + \Delta\varphi_-)$$

For phase modulation ($\Delta\varphi_- = \Delta\varphi_+ + \pi$), we have

**Eq. (S11)** $$I_{det\_RF} \propto 2\gamma_{PD}[E_c E_- \cos(\omega_{RF} t + \Delta\varphi_-) + E_c E_- \cos(\omega_{RF} t + \Delta\varphi_- + \pi)]$$
$$= 2\gamma_{PD}[E_c E_- \cos(\omega_{RF} t + \Delta\varphi_-) - E_c E_- \cos(\omega_{RF} t + \Delta\varphi_-)] = 0$$

It is clear that the detected photocurrent reaches its maximum at intensity modulation, and when the two sidebands are out of phase, there is no RF current. Additionally, even when the two sidebands have no phase difference, the phase difference between sidebands and carrier will influence the value of photocurrent. When all of the three fields carry the same phase value, the photocurrent is the highest.

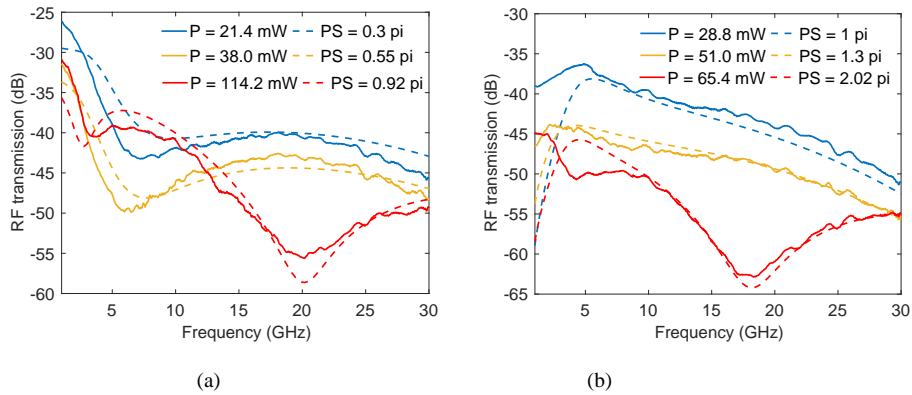

Fig. S3 Simulated (Dashed-line) and measured (Solid-line) results of RF transmission based on intensity modulated signal as input. (In the figure legend 'P' stands for the power applied to the phase shifter and 'PS' is the phase shift we set in the simulation.)

## 2b. Modelling of modulation transformation experiments

In Figure 4 of the paper we give the calculated and experimental results of modulation transformation experiments from 15 to 25 GHz, and here we show the results in a full spectral range (1 to 30 GHz). Fig. S3 (a) and (b) show the IM to PM and PM to IM conversion respectively. According to the heater parameters in our silicon chip, we need a power around 35 mW to create a 'π' phase shift. In Fig. S3 (a), from the IM to PM (blue to red line), the power difference is 114.2-21.4 = 92.8 W, which divided by 35W is 2.65(π), means the induced phase shift is around 0.65 π, that matches the phase shift in the simulation (0.62 π). Also, the RF loss in the system will shift the peak of the curves. The peak of the blue curve in Fig. S3 (a) is supposed to be at 25 GHz in a no-loss case. This tells us why the phase shift is not π but 0.65 π. For the PM to IM conversion experiment, due to the red and orange lines are relatively linear, the RF loss will not change the trend of these curves. A π phase shift switched PM to IM in both simulation and experiment.

## 3. Results of amplitude tuning

The modulation transformation experiments confirm the flexibility of phase tuning in the spectrum synthesis circuit, where we mainly tuned the phase shifter and left the tunable coupler at a fixed stage. To verify the versatility of the circuit, we also performed amplitude tuning experiments where we vary the voltage applied on the tunable coupler to control the sideband attenuation.

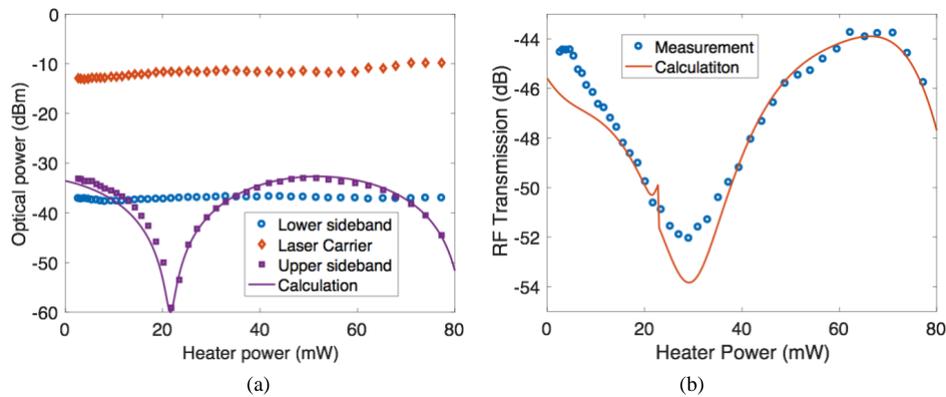

Fig. S4 Results of amplitude tuning experiment: (a) Optical power of three light waves; (b) Corresponding RF transmission.

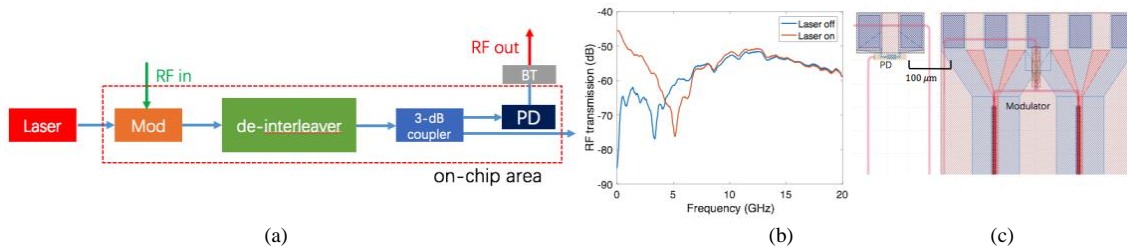

Fig. S5 Experimental result for on-chip link performance: (a) Experiment schematic; (b) Measured response when the laser is on and off. (c) Distance between PD and modulator in our silicon chip layout.

The experiment setup is same as the modulation transformation experiment. The input is a phase modulated signal. We gradually tuned the applied voltage on the phase shifter attached on one of the arm of the tunable coupler. A single frequency point (here 20 GHz) is selected to record the optical spectrum and RF transmission.

Fig. S4 (a) shows the intensity of three light waves in the amplitude tuning experiment. These data points are taken from each optical spectrum at different applied voltages. The applied voltages are transferred to heater power using Ohm's law, which constitutes the x-axis in Fig. S4. The amplitude (power) of lower sideband and laser carrier were fixed, while the amplitude of the upper sideband was varied. Fig. S4 (c) shows the corresponding RF transmission data and theoretical calculation. A slight shift in the heater power corresponding to the minima of the RF transmission with respect to the minima in the sideband power is attributed to the phase rotation from tuning the tunable coupler (see Supplementary 1b).

## 4. Internal RF crosstalk

There are mainly two reasons that the on-chip modulator is not applied to the experiment: bandwidth limitation and internal RF crosstalk. To characterize the impact of crosstalk, a complete experimental on-chip RF photonic link is established, as is shown in Fig. S5 (a). The measured $S_{21}$ traces are shown in Fig. S5 (b). The result shows, with laser turned on or off, the measured responses merely show the difference in the frequency regime that is lower than 5 GHz.

The explanation for this behaviour lies in the existence of internal RF crosstalk, where the distance between the electrodes of modulator and PD are very close (Fig. S5 (c)), thus even there is no optical signal, a significant RF signal will transmit from the input pins of the RF needle of the modulator directly to the pins of PD through the chip surface and free space. When the laser is turned on, the direct transmit signal will interfere with the photocurrent. A reduction of the internal crosswalk would require a larger space between modulator and PD, of at least 1-2mm.

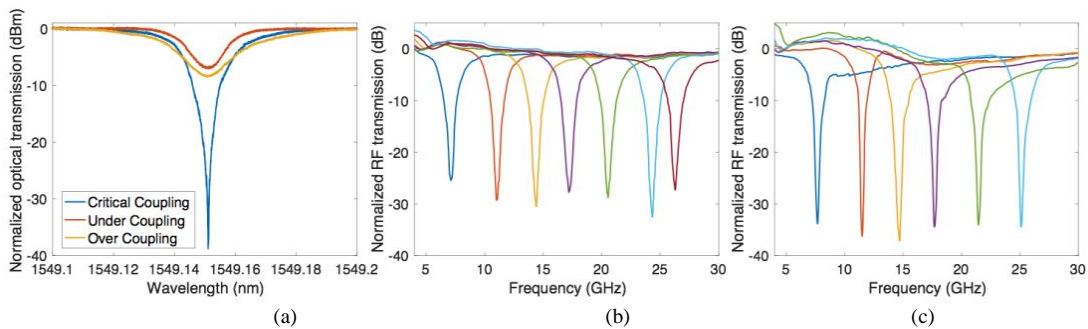

Fig. S6 Experimental results at Different coupling status of all-pass ring; (b) Tunability of SSB notch filter; (c) Tunability of phase cancellation notch filter.

## 5. Thermal tuning of all-pass ring and filters

There are two heaters attached to the all-pass ring that control the coupling and the phase (resonance) respectively. By changing the coupling coefficient of the ring we can realize different coupling status from over to under coupling. Fig. S6 (a) shows different coupling state of the ring with thermal tuning. The tunability of our SSB notch filter and phase cancellation filter are shown in Fig. S6 (b) and (c). For the SSB notch filter the ring is at critical coupling.